\documentstyle [11pt]{article}

\newbox\grsign \setbox\grsign=\hbox{$>$} \newdimen\grdimen
\grdimen=\ht\grsign
\newbox\simlessbox \newbox\simgreatbox \newbox\simpropbox
\setbox\simgreatbox=\hbox{\raise.5ex\hbox{$>$}\llap
     {\lower.5ex\hbox{$\sim$}}}\ht1=\grdimen\dp1=0pt
\setbox\simlessbox=\hbox{\raise.5ex\hbox{$<$}\llap
     {\lower.5ex\hbox{$\sim$}}}\ht2=\grdimen\dp2=0pt
\setbox\simpropbox=\hbox{\raise.5ex\hbox{$\propto$}\llap
     {\lower.5ex\hbox{$\sim$}}}\ht2=\grdimen\dp2=0pt

\title{\vspace{4cm}\Large\bf
Eliminating the `flatness problem'\\
with the use of Type Ia supernova data
}
\author {\bf A.D.~Chernin\\
{\it Sternberg Astronomical Institute, Moscow University, Moscow, 119899,
Russia}\\
{\it Tuorla Observatory, Turku University, Piikki\"o, 21500, Finland},\\
{\it Astronomy Division, Oulu University, Oulu, 90014, Finland}
}
\date{~}
\begin{document}
\maketitle
\begin{abstract}
\noindent
{\bf
Recent measurements of the cosmological constant or cosmic vacuum in Type Ia SN observations
(Riess et al.1998, Perlmutter et al. 1999) imply that $\Omega (t)$ is
exactly unity or  nearly unity at any epoch of cosmic evolution.
No fine tuning is needed to describe this phenomenon with the standard  cosmological model based
on the Type Ia SN data. A time-independent symmetry relation  between vacuum and matter is
basically behind the observed near-parabolic expansion.
\vspace{0.5cm}
\noindent

}
\end{abstract}
\vspace{0.5cm}
\newpage

\section{Introduction}

It was first mentioned long ago (Dicke 1970), that the universe must be `extremely
finely tuned' to yield the present observed balance between the kinetic energy
of expansion $K$ and the gravitational potential energy of cosmic matter $U$. The balance is
usually quantified in terms of $\Omega$, which is the energy ratio, $\vert U \vert /K$,
and also the ratio
of the  total matter density, $\rho$, to the critical density, $\rho_c$. Thirty years ago,
observational limits on $\Omega$ were described
as $0.1 < \Omega_0 < 10$, and it was argued that such an apparently wide range implies a
very narrow range at earlier epochs. It has been estimated more than once since that
 $\Omega$  departs from unity by one part in $10^{16}$ at the
epoch of light element production, or by one part in
$10^{60}$ at the Planck epoch. Why was there such a remarkably fine balance between the
kinetic and potential energies in the `initial conditions' for the cosmic evolution?

This argument has become known as the `flatness problem',
since the energy ratio is associated with the sign of the spatial curvature in the Friedmann model.
It serves as one of the main
motivations for the idea of inflation (Guth 1981); this idea suggests an
elegant solution to the problem (Guth 1981, Linde 1990). Various aspects of
the problem have discussed by Frieman \& Waga (1998), Kaloper et al. (1999),
Berera et al. (1999),
Albrecht \& Magueijo (1999), Barrow \& Magueijo (1999), Kaloper \& Linde (1999), Clayton \&
Moffat (1999), Avelino \& Martins (1999), Arkani-Hamed et al. (2000), Pierpaoli et al. (2000),
Kavasaki et al. (2000), Roos \& Harun-or-Rashid (2000), Kaganovich (2001), Tegmark et al.
(2001), Barrow \& Kodama (2001), Starkman et al. (2001), Youm (2001), Deffayet et al.
(2001), to mention only some publications of the last 2-3 years.

In this {\it Letter}, I show that the recent Type Ia supernova measurements  (Riess et al. 1998,
Perlmutter et al. 1999) displays the fine-tuning argument in a new light.
In Sec.2, a brief account of the concordant data and the new standard model which is based on them
is given; new general constraints on $\vert \Omega (t) -1 \vert$ for any epoch of the cosmic evolution
are presented in Sec.3; the results are discussed in Sec.4.

\section{Standard model re-visited}

The dynamical equation of the Friedmann cosmology  has a form of energy conservation
($K + U = Const$):
\begin{equation}
{\dot a}^2 = (a/A_V)^2 + (a/A_D)^{-1} + (a/A_B) + (a/A_R)^{-2} - k,
\end{equation}
where $a(t)$ is the curvature radius and/or scale factor of the model; $k = +1, 0, -1$
for elliptic  ($K+U < 0$) expansion with positive spatial curvature, parabolic  expansion
($K+U = 0$) with zero curvature and hyperbolic  expansion ($K + U > 0$) with negative
curvature, respectively.

The constants $A$ in Eq.(1) represent the four major cosmic energy components
which are vacuum (V), dark matter (D), baryons (B) and radiation or relativistic energy (R).
 The constants are the integrals of the Friedmann `thermodynamical' equation:
\begin{equation}
A = [\frac{8 \pi G}{3} \rho a^{3(1+w)}]^{\frac{1}{1+3w}},
\end{equation}
where $w$ is the pressure-to-density ratio which is $-1, 0, 0, 1/3$ for vacuum, dark matter,
baryons and radiation, respectively. Hereafter, the integrals $A$ are called the Friedmann
constants. The constants are related to the `initial condition', under which the
cosmic energy forms, vacuum including, were generated in the early universe.

The Friedmann constants can be evaluated quantitatively with the use of the current
concordant figures for the four energy densities (Riess et al.1998, Perlmutter et al. 1999,
see also for a review Primack 2000):

\begin{equation}
\Omega_V = 0.7 \pm 0.1,\; \Omega_D = 0.3 \pm 0.1,\; \Omega_B = 0.02 h_{100}^{-2}, \;
\Omega_R = 0.6 \alpha \times 10^{-4}, \; 1 < \alpha < 10-30.
\end{equation}
\noindent
These figures are in general agreement with the Hubble constant
$h_{100} = 0.70 \pm 0.15$ and the cosmic age $t_0 = 15 \pm 3$ Gyr. More rigorous
recently published upper and lower  bounds on the cosmic age are
$13.2 {+1.2\atop -0.8}$ Gyr for $h_{100} = 0.72 \pm 0.08$ (Ferreras et al. 2001).

The present-day value of $a(t)$ is determined by these figures; approximately,
on the order of magnitude, one has: $a_0 = a(t_0) \sim A_V$, for all the three types of expansion
and the signs of spatial curvature.

The four Friedmann constants  calculated with the concordant data are proved to be coincident,
on the order of magnitude (Chernin 2001):
\begin{equation}
A_V \sim A_D \sim A_B \sim A_R \sim A \sim 10^{60 \pm 1} M_{Pl}^{-1}.
\end{equation}
Here,  units are used in which the speed of light, the Boltzmann constant and the Planck constant
are all equal unity: $c = k = \hbar = 1$. The Planck mass is
$M_{Pl} = G^{-1/2} \simeq 1.2 \times 10^{19}$ GeV.

The time-independent coincidence of the Friedmann constants described by Eq.(4) looks like
 a symmetry relation that brings vacuum into association with non-vacuum cosmic
energies. (A similar relation for baryons and radiation was recognized soon after the
discovery of the CMB -- Chernin 1968).
As I show below, the symmetry relation between vacuum and matter puts a robust upper
bound on $\vert \Omega (t) -1 \vert $ at any epoch of cosmic evolution.

\section{Setting constraints on $\vert \Omega (t) -1 \vert$}

The concordant data of Sec.2 narrow essentially the observational bounds on
$\Omega (t_0) = \Omega_0$,
in comparison with what was discussed three decades ago (see Sec.1). A conservative restriction
can be seen from Eq.(3) above: $0.8 < \Omega_0 < 1.2$.
More stringent limits (like $\Omega_0 = 0.97 \pm 0.05$) are also advocated in current
literature (Roos \& Harun-or-Rashid 2001). From the point of view of
the fine-tuning argument, the new data make the problem
harder than it was in 1970, but not too much -- it was hard enough from the very beginning.
It is more significant that the Type Ia supernova measurements (Ries et al.1998, Perlmutter et al.
1999) have changed the very sense of the problem.

In the new standard model described in Sec.2, the dynamics of expansion  exhibits the parabolic
($K + U = 0$)
behaviour as $t$ goes to both zero and infinity for flat, closed and open models, as is seen from
Eq.(1).
Considerable (maximal) deviations from the regime can be expected for $k \ne 0$  in an
`intermediate' cosmic time interval.

As for spatial geometry, the ratio of the horizon radius to the radius of 3-curvature, $ct/a$,
may be a practical measure of non-flatness. This ratio goes to
zero when time goes to both zero and infinity, as is also seen from Eq.(1).
It means that non-flatness wanishes at
early and late epoch of the cosmic evolution. In both closed and open models,
the radius of 3-curvature is comparable with the radius of the horizon at the intermediate times.
In this sense, non-flatness is finite and, generally, not small in
the intermediate time interval, including the present epoch at which
$a \sim A_V \sim ct$ (see. Sec.2).

To put these considerations in a quantitative way, let us follow the evolution
of $\Omega$ as a function of time.
It is easy to see from Eq.(1) that
\begin{equation}
\Omega (t) = [1 - k ( \frac{8 \pi G}{3} \rho (t) a(t)^2)^{-1}]^{-1},
\end{equation}
\noindent
where $\rho (t)$ is the total density:
\begin{equation}
\frac{8 \pi G}{3} \rho = A_V^{-2} + A_D a^{-3} + A_B a^{-3} + A_R^2 a^{-4}.
\end{equation}

In the limits $t \rightarrow 0$ and $t \rightarrow \infty$, the function $\Omega (t)$
reaches the  unity level, and the maximal deviation from unity may be expected in the
intermediate time interval mentioned above.
In this time interval (starting, say, with $ a  \ge (0.01 - 0.1) a_0$),
the most significant  contributions to the total density
$\rho$ are provided by the vacuum density $\rho_V $ and the dark matter density
$\rho_D$. Thus, with good accuracy, one can re-write Eq.(5) in the simple form:
\begin{equation}
\Omega (t) \simeq [1 - k ( \frac{a^2}{A_V^2} +  \frac{A_D}{a})^{-1}]^{-1}.
\end{equation}

At $a = a_{ex} = (\frac{1}{2} A_V^{2} A_D)^{1/3}$ the function $\Omega (t)$ has an extremum,
\begin{equation}
\Omega_{ex} \simeq [1 - \frac{1}{2} k ( \frac{A_D}{A_V})^{2/3}]^{-1}.
\end{equation}
\noindent
Putting in accordance with Eq.(4) $A_V \simeq A_D$, one finds that there are
a minimum $\Omega_{ex} \simeq 2/3$ for $k = -1$ and a maximum $\Omega_{ex} \simeq 2$ for $k = 1$.

This leads to upper bounds  on deviation from the unity level of $\Omega (t)$ valid for all
the times of the cosmic evolution:
\begin{equation}
 \Omega (t)- 1 \le 1, \;\;\; k = 1; \;\;\;\;\;\; \vert \Omega (t) - 1 \vert \le 1/3,
\;\;\;\;\;\; k = -1.
\end{equation}

Defined and estimated in this way, the upper bounds of Eq.(8) prove to be rather
severe. They rule out completely any possibility of large deviations from the near-parabolic
dynamics at any epoch of cosmic expansion. Numerically, the bounds are not too different from
the present-day observed figures; the latter could be expected,  because the epoch of extremum,
$ a_{ex} \simeq 2^{-1/3} A_V$, is not far from us in look-back time.

\section{Dicussion}

The new standard model inspired by the Type Ia supernova measurements  (Riess et al.1998,
Perlmutter et al. 1999) describes not only the present transition epoch of the cosmic
expansion, but as well the early epoch of matter (radiation) domination and the later
epoch of vacuum domination (Sec.2). The model leads to the strong conclusion
that the cosmic expansion is exactly parabolic or  nearly parabolic during all the
cosmic evolution. Eqs.(5-8) show this in an explicit
quantitative way and clarify also the basic cause of the nearly-parabolic dynamics in the
standard model.

The physics which is behind the phenomenon is directly related to the Friedmann constants
of the
standard model. According to Eq.(7), the constraints on possible deviations from
the parabolic dynamics are expressed in the terms of the Friedmann constants. The (practically
exact) relation of Eq.(7) includes the  vacuum
constant $A_V$ and the dark matter constant $A_D$. The Type Ia SN data put the constants in
a simple relation: $A_V \sim A_D$. This approximate relation is
enough to provide the upper bound on possible deviations from the parabolic regime of the
cosmic expansion for all the cosmic time.

The quantitative estimate of Eqs.(7,8) and the qualitative conclusion that follows from it
are  obviously stable and robust: they are insensitive to small variations of the
constants involved. Moreover, the estimate does not depend on the absolute values of the
individual constants; only the constant ratio $A_D/A_V$, which is unity or near unity,
affects the result. In this sense, the presently observed near-parabolic expansion is
completely controlled by only one order-of-unity empirical parameter of the model.

Thus, the long-standing problem of the near-parabolic dynamics finds a clear solution: the
cosmic expansion is basically as it is because the Friedmann constants are nearly coincident.

There is no need in addressing a hypothetical pre-Friedmann inflation to solve the problem.
As for the fine-tuning argument (see Sec.1), it appears now in a
quite different light. As is seen from Eqs.(5-7), the seemingly fine-tuned inter-relation
between the time-dependent kinetic $K$ and potential $U$ energy at an early epoch is
robustly determined by the time-independent symmetry relation of Eq.(4).

Instead of the `fine tuning' argument, one faces now a new question: Why are the
Friedmann constants nearly coincident?

The physical nature of the Friedmann constants
seems to be associated with the origin of the cosmic energy forms in the very early
universe. One may expect that the physical processes that developed at that time
might bring the four cosmic energies into a correspondence with each other.
A possible approach to the problem  (Chernin 2001) assumes that the freeze-out process at
the epoch of the electroweak ($\sim 1$ TeV) temperatures might be
responsible for establishing the symmetry relation between vacuum and matter.
If so, the `initial conditions' for the observed cosmic dynamics were generated in
the terms of the Friedmann constants at cosmic age $t \sim 10^{-12}$ sec.

\vspace{0.3cm}
I thank Dr. R. D\"ummler of Oulu University for helpful comments.

\newpage

\section*{References}

Albrecht A., Magueijo J.,  1999, Phys. Rev., D59, 043516

Arkani-Hamed N., Dimopoulos S., Kaloper N., Sundrum R., 2000,
Phys. Lett., B480, 193

Avelino P.P., Martins C.J.A.P., 1999, Phys. Lett., B459, 46

Barrow J.D., Magueijo J., 1999, Phys. Lett., B447, 24

Barrow J.D., Kodama H., 2001, Class. Quant. Grav., 18, 1753

Berera A., Gleiser M., Ramos R.O., 1999, Phys. Rev. Lett., 83, 264

Chernin A.D., 1968, Nature, 220, 250

Chernin A.D., 2001, Physics-Uspekhi, 71, 1153 (preprint: astro-ph/0110003)

Clayton M.A., Moffat J.W., 1999, Phys. Lett., B460, 263

Deffayet C., Dvali G., Gabadadze G., Lue A., 2001, Phys. Rev., D64, 104002

Dicke R.H., 1970, Gravitation and the Universe. Amer. Phil. Soc., Philadelphia

Frieman J.A., Waga I., 1998, Phys. Rev., D57, 4642

Guth A., 1981, Phys. Rev., D23, 347

Kaganovich A.B., 2001, Phys. Rev., D63, 025022

Kaloper N., Linde A., 1999, Phys. Rev. D59, 10130

Kaloper N., Linde A., Bousso, R., 1999, Phys. Rev., D59, 043508

Kavasaki M., Yamaguchi M., Yanagida T., 2000, Phys. Rev. Lett., 85, 3572

Linde A., 1990, Particle physics and inflationary cosmology. AIP, New York

Perlmutter S. et al., 1999, ApJ, 517, 565

Pierpaoli E., Scott D., White M., 2000, Mod. Phys. Lett., A15, 1357

Primack J.R., 2001, Nucl. Phys. B Proc. Suppl. 87, 3

Riess A.G. et al., 1998, AJ, 116, 1009

Roos M., Harun-or-Rashid S.M., 2000, preprint (astro-ph/0005541)

Starkman G.D., Stojkovic D., Trodden M., 2001, Phys. Rev., D63, 103511

Starkman G.D., Stojkovic D., Trodden M., 2001, preprint (astro-ph 0106143)

Tegmark M., Zaldarriaga M., Hamilton A.J.S., 2001, Phys. Rev., D63, 043007

Youm D., 2001, Phys. Rev. D63, 125011

\end{document}